\documentclass[12pt]{article}

\usepackage{amsmath}

\begin{document}

\begin{center}
{\bf
Red'kov  V.A.,  George J. Spix \\[2mm]
On  the different forms of the Maxwell's electromagnetic
\\equations
in a uniform media}\\[2mm]
 Institute of Physics, National  Academy of Sciences
of Belarus\footnote{redkov@dragon.bas-net.by }\\
Bachelor of Science in Electrical Engineering,  Illinois Institute
of Technology\footnote{gjspix@msn.com}
\end{center}

\begin{quotation}
Two known, alternative to each other, forms of the
Maxwell's electromagnetic equations in a moving uniform media  are
investigated and discussed. Approach commonly used after Minkowski
is based on the two tensors:
 $H^{ab} = ({\bf D}, {\bf H} /c)$ and  $F^{ab} = ({\bf E}, c {\bf B}) $ which  transform independently
 of each  other at Lorentz transitions;
relationships between fields ${\bf D} = \epsilon_{0}\epsilon {\bf
E}, {\bf B} = \mu_{0}\mu {\bf H}$ change their form  at Lorentz
transformations and have the form of the Minkowski equations
depending on      the  4-velocity $u^{a}$  of the moving media
under an inertial reference frame. In this approach, the wave
equation for electromagnetic potential involve explicitly the
$u^{a}$-velocity of the moving media. So, the electrodynamics by
Minkowski implies the absolute nature of the mechanical motion. An
alternative formalism (Rosen and others)   may be developed in the
new variables: $ c = 1 /\sqrt{\epsilon_{0} \mu_{0}} , \; k
=1/\sqrt{\epsilon \mu}, \;  x^{0} = kct, \; j^{0} = J^{0}, \; {\bf
j} =  {\bf J} / kc ,;{\bf d} = \epsilon_{0} \mu_{0}, \; {\bf E}
,\;  {\bf h} = {\bf H } / kc $.
 In these  variables, the  Maxwell's equations can be written  in terms of
 a single  tensor  $f^{BC} =({\bf d}, {\bf h})$.
This form of the  the Maxwell's equations exhibits symmetry under
modified Lorentz transformations in which  everywhere instead of
the vacuum speed of light  $c$ is used the speed of light in the
media,
 $kc $. In virtue of this symmetry we might consider
 such a formulation of the Maxwell theory in the media as
 invariant under the  mechanical motion of the reference frame.
In connection with these two theoretical schemes,  a   point  of
principle  must be stressed: it might seem well-taken the
requirement to perform Poincar\'{e}-Einstein clock synchronization
in the
 uniform  medias
with the help of  real light signals influenced by the  media,
which  leads us to the modified Lorentz symmetry.


 \end{quotation}

\section{ Maxwell equations in a media, transition to new
variables }

\hspace{5mm}
Maxwell's equations in a uniform media with two characteristics
 $\epsilon >1$ and $\mu>1$ (dielectric and magnetic penetrability) have the
 form [1]
\begin{eqnarray}
 \mbox{div} \; {\bf D} =  \; \rho \; , \qquad
\mbox{div} \;  {\bf B} = 0 \; , \qquad
  \mbox{rot} \;{\bf H}  =
{\bf J} +    {\partial{\bf D} \over \partial t} \; ,\qquad
\mbox{rot}\; {\bf E}  = -  {\partial  {\bf B}\over
\partial t}    \; .
\label{5.1.1d}
\end{eqnarray}

\noindent
The notation is used:
\begin{eqnarray}
\bf B       = \mu  \mu_{0} \; {\bf H} \; , \qquad \epsilon
\epsilon_{0} \; {\bf E} = {\bf D} \; , \qquad c = {1 \over
\sqrt{\epsilon_{0} \mu_{0}}  } \; , \qquad k = {1 \over
\sqrt{\epsilon \mu} } < 1 \; . \label{5.1.4}
\end{eqnarray}

The  speed of light in  the  media $c_{media}$   is less than that in vacuum
  $c$ and the coefficient  $k$ describes  this decreasing:
$ c_{media} = k \; c $.
Now a principal point in further analysis is that  eqs. (\ref{5.1.1d}) may be re-written
as
\begin{eqnarray}
 \mbox{div} \; {\bf D} =   \rho \; ,\qquad
\mbox{div} \; { {\bf H} \over kc}  = 0 \; , \qquad
\mbox{rot}\;  {{\bf H} \over kc}   =   {{\bf J} \over kc } \; +\;
{\partial \over \partial (kct)} \;  {\bf D}   \; ,
\qquad
 \mbox{rot}\; {\bf D}  = - \;  {\partial \over
\partial ( kct )} \; {{\bf H} \over kc}   \; .
\label{5.1.7}
\end{eqnarray}

\noindent
Instead of variables
$(t,  x^{i}) ,  (\rho,   {\bf J}),   ({\bf D},  {\bf H}) ,
$
you may define new   ones by means of the formulas
\begin{eqnarray}
x^{0} = kc \; t  , \qquad  j^{0} = \rho , \;\;
{\bf j} = {{\bf J} \over kc} , \qquad   {\bf d} = {\bf D}, \;
{\bf h} = {{\bf H} \over kc} \; .
\label{5.1.8}
\end{eqnarray}

\noindent
On making so, Maxwell's equations (\ref{5.1.7}) will take the form
\begin{eqnarray}
 \mbox{div} \; {\bf d} =   j^{0} \; ,\qquad
\mbox{div} \;  {\bf h}   = 0 \; , \qquad
\mbox{rot}\;  {\bf h}   =   {\bf j}  \; +\;
{\partial \over \partial x^{0}} \;  {\bf d}   \; , \qquad
\mbox{rot}\; {\bf d}  = - \;  {\partial \over
\partial x^{0} } \;  {\bf h}    \; .
\label{5.1.9}
\end{eqnarray}

\section{ Maxwell's equations  and  Lorentz transformations  in the  media }

\hspace{5mm}
 Maxwell's equations  (\ref{5.1.9}) can be written down in
the explicit comprehensive form:
\begin{eqnarray}
\partial _{1} \; d^{1}  +
\partial _{2} \; d^{2}  + \partial_{3} \; d^{3}   =   j^{\;0} \; , \qquad
\partial _{1} \; h^{1}  +
\partial _{2} \; h^{2}  + \partial_{3} \; h^{3}   = 0 \; ,
\nonumber
\\
\partial _{2}  h^{3} - \partial _{3}  h^{2} = j^{1} + \partial_{0} d^{1} ,\qquad
 \partial _{3}  h^{1} - \partial _{1} h^{3} = j^{2} + \partial_{0} d^{2}\; ,
 \qquad
\partial _{1}   h^{2} - \partial _{2}   h^{1} = j^{1} + \partial_{0} d^{1}\; ,
\nonumber
\\
\partial _{2}  d^{3} - \partial _{3} d^{2} = -  \partial_{0} h^{1} ,
\qquad
\partial _{3}  d^{1} - \partial _{1}  d^{3} = -   \partial_{0} h^{2} , \qquad
\qquad
\partial _{1}  d^{2} - \partial _{2}   d^{1} = -  \partial_{0} h^{1}  .
\label{5.1.14}
\end{eqnarray}

\noindent
Now let us introduce certain linear transformation
over quantities entering  the  Maxwell's equations (modified Lorentz transformation):
\begin{eqnarray}
x^{\;'0} = \mbox{ch}  \sigma  \; x^{\;0} - \mbox{sh}\sigma \; x^{1} \; \qquad
x^{\;'1} = - \mbox{sh}  \sigma  \; x^{\;0} + \mbox{ch}\sigma \; x^{1} \; , \qquad
x^{\;'2} = x^{2}\; , \; x^{\;'3} = x^{3}\; ,
\nonumber
\\
j^{\;'0} = \mbox{ch}  \sigma  \; j^{\;0} - \mbox{sh}\sigma \; j^{1} \; ,\qquad
j^{\;'1} = -\mbox{sh}  \sigma  \; j^{\;0} + \mbox{ch}\sigma \; j^{1} \; ,\qquad
j^{\;'2} = j^{2}\; , \; j^{\;'3} = j^{3}\; ,
\nonumber
\\
d^{\;'1} = + d^{1} \; ,   \qquad
d^{\;'2}= \mbox{ch}  \sigma  \; d^{2} - \mbox{sh}  \sigma  \; h^{3} \; , \qquad
h^{\;'3}=    -\mbox{sh}  \sigma  \; d^{2} +\mbox{ch}  \sigma  \; h^{3}  \; ,
\nonumber
\\
h^{\;'1} = + h^{1} \; ,   \qquad
d^{\;'3}= \mbox{ch}  \sigma  \; d^{3} + \mbox{sh}  \sigma  \; h^{2} \; , \qquad
h^{\;'2}=    +\mbox{sh}  \sigma  \; d^{3} + \mbox{ch}  \sigma  \; h^{2}  \; .
\label{5.1.15}
\end{eqnarray}

Now the  task is to show that if one transforms
equations (\ref{5.1.15}) to new (primed) variables,
then as a result one again will obtain equations
of the form  (\ref{5.1.14}) with a single difference:
 all quantities become  primed ones (see also [11-14]):
 \begin{eqnarray}
 \mbox{div} '\; {\bf d}' =   j^{\; '0}  \; ,\qquad
\mbox{div}' \;  {\bf h} '  = 0 \; , \qquad
\mbox{rot}'\;  {\bf h}'   =   {\bf j}'  \; +\;
{\partial \over \partial x^{'0}} \;  {\bf d}'   \; ,
\qquad
 \mbox{rot}'\; {\bf d}'  = - \;  {\partial \over
\partial x^{'0} } \;  {\bf h} '   \; .
\label{5.1.30}
\end{eqnarray}

\section{On physical sense of the modified Lorentz transformations}

\hspace{5mm}
What is the physical sense of the dimensionless parameter  $\sigma$ in the above
Lorentz formulas?
 In fact, from the very beginning the question was:
how Maxwell's equations behave themselves when   reference frame is changed from
$K$ to a moving  $K'$.
For the  situation when velocity is small enough we must obtain
a simple and "evident" \hspace{2mm} solution in the form of
 Galilei  formula for a  coordinate transform:
\begin{eqnarray}
t'= t, \; x' = x - Vt, \; y'=y, \; z'=z \; .
\label{5.1.33}
\end{eqnarray}

The Lorentz transformation at small $\sigma$  will take the form
\begin{eqnarray}
kc\; t '  \approx   \; kct  - \sigma \; x  \;  \approx \; kct  \;:
\; \Longrightarrow \; t'=t \; ,\;
%
x '  =  - \sigma  \; kc\; t  +  x =  x - Vt \; , \; \mbox{if}
\;\;  \sigma =   {V \over kc}\; .
\nonumber
\end{eqnarray}

So that the physical sense of the parameter
$\sigma$ (at its small values  $\sigma <<1$ ) is found:
\begin{eqnarray}
\sigma << 1: \qquad \Longrightarrow \qquad
\sigma =   {V \over kc}  =  { V \over c_{media}} \; .
\label{5.1.37a}
\end{eqnarray}

\noindent
One needs to  \underline{generalize} eq.  (\ref{5.1.37a}) for arbitrary values of $V$.
This is achieved by the following relations:
\begin{eqnarray}
0< \mid V \mid < kc \; \; : \qquad
\mbox{ch}\; \sigma = {1 \over \sqrt{1 - (V/kc)^{2}}}\; , \;\;
\mbox{sh}\; \sigma =  {(V /kc)  \over \sqrt{1 - (V/kc)^{2}}}\; . \qquad
\label{5.1.37b}
\end{eqnarray}

\noindent or
\begin{eqnarray}
t' = {  t - V  x /k^{2}c^{2}  \over \sqrt{1 - (V/kc)^{2}}} \;,\qquad
x' = {  \;x -Vt \over \sqrt{1 - (V/kc)^{2}}} \; .
\label{5.1.38a}
\end{eqnarray}

\section{ The speed of light  and the  modified Lorentz transformations}

\hspace{5mm}
On finding the modified Lorentz transformations a simple kinematical problem of
modified  Lorentz formulas for velocity may be  immediately solved.
It provides us with a postulate on light velocity  ($kc$) constancy, the  crucial  logical  element
in Einstein's construction of Special Relativity [4].

So it may be readily derived the modified rule for  transforming the velocity vector
${\bf W}=(W_{x},W_{y},W_{z})$:
\begin{eqnarray}
W_{\;'x}  = {  W_{x} - V   \over 1 -    V W_{x} / k^{2} c^{2} }  \; ,
\;\;
W_{\;'y} =  {\sqrt{1 - V^{2} / k^{2} c^{\;2} } \over
1 -   V W_{x} / k^{2} c^{2}  } \;\; W_{y} \; ,
\;\;
W_{\;'z} =  {\sqrt{1 - V^{2} / k^{2} c^{\;2} } \over
1 -   V W_{x} / k^{2} c^{2}  } \;\; W_{z} \;
\label{5.1.43}
\end{eqnarray}

\noindent This is  a  modified version of the  famous rule for velocity summing by
Lorentz-Poincar\'{e}-Einstein. This formulas results in some consequences.

\begin{quotation}

{\em \noindent Lorentz transformation along the axis  $x$ does not
change the value of the speed of  light propagating along this
direction $x$. Lorentz transformation along the axis $x$ does not
change the modulus of  the light velocity vector.}

\end{quotation}

\section{Lorentz-Poincar\'{e}-Einstein, controversy and misunderstanding}

\hspace{5mm}
It was Lorentz  [2] who first established a remarkable  property of the
Maxwell's equations: its symmetry under special mathematical transformations when
entering these equations quantities -- time and space coordinates, charge-current density, and
electromag\-netic fields. Poincar\'{e} introduced clarity  [3] into Lorentz  initial
 formulas and revealed its mathematical (so-called) group structure.
Undoubtedly the first deciding steps on the road to Special relativity  theory
were made by Lorentz and this was stressed by Poincar\'{e}  more than once.
In the same time, Lorentz never
ascribes to himself everything on  this road and   very highly appreciated
the role and contribution  of Poincar\'{e}.

Unfortunately, afterwards in connection with Special relativity there arose  controversy
and  misunderstanding on the question -- who is creator of this theory:
Lorentz, Poincar\'{e},  or Einstein.  To the present day this dispute is still with us
(reviewing of the situation, for example, see in [15]).
We will not join the
debaters. In our opinion, all three, Lorentz, Poincar\'{e}, and Einstein,
are creators of the theory\footnote{Also we should not ignore a great number of physicists,
owing to their   enormous and  laborious work we could go up mounts
where only three great names are present.}.

The first was Lorentz, then Poincar\'{e} sided with  him, and next
Einstein  started his  work on creating Special relativity [4], mainly on
its physical interpretation,  comprehension,   and logical reconstruction.
The question -- who is the \underline{main}  creator -- is false.
Lorentz formulated his view on the matter concisely and  definitely  [5]:
{\em
the same what we had deduced from Maxwell's equations Einstein has postulated
...}.

So, creating  Special relativity  has proceeded along two lines,
and it is absurd to consider them as
absolutely independent. One  line goes upward to Special relativity  from
symmetry property of Maxwell's theory in moving bodies. This is \underline{inductive}
way and it is historically first one. The second   line,
though logically independent in appearance, is  \underline{deductive}
construction of the theory by going down from a special postulate [4]. But the postulate
itself  can be regarded as a  logical
mathematical result of the Lorentz-Poincar\'{e} analysis of  the Maxwell's theory.

Logical treatment suggested by Einstein seems for many simple and  clear, so that it may be
explained even to a person without any special education.
This circumstance   assists in the promotion of Einstein treatment of
Special relativity and its notability in general public.

However by creating  the deductive  way to construct Special relativity
does not  provide  the ground for assigning to A. Einstein the main or say  single
creator of the  theory.
Lorentz and Poincar\'{e} provided us with inductive  way to this theory, and this
way was historically the first. Both lines to Special relativity are legitimate
and  mutually complementary.

\section{ On the  Maxwell theory  in
 the  media,  Minkowski approach
  }

\hspace{5mm}
Although the  main work on Special relativity of A. Einstein [4] in 1905  is titled
 "On  the Electrody\-na\-mics of
Moving Bodies", in this paper Maxwell equations  only in vacuum
(a media with trivial values  $\epsilon=1, \mu =1$ ) had been
considered   in fact and the  symmetry properties of these equations had been used.
In accordance  with this  in all theoretical building
from the very beginning only a universal light velocity in vacuum was used and just for  that  velocity
was advanced a postulate of its constancy irrespective of the  motion  of
 the  reference frame.

Later in 1908 H. Minkowski  gave [6]  a more detailed and accurate treatment
of the Maxwell theory in a  uniform media $(\epsilon \neq 1, \mu \neq 1)$
with respect to requirements of Special relativity.
Two points of   his study should be emphasized:

{\em Minkowski had elaborated a very convenient and still actively exploited mathematical
technique
-- so called 4-dimensional tensor formalism\footnote{To be exact, H.Poincar\'{e}
had proposed and developed in some  aspects the same technique before Minkowski [4].}.

Minkowski had found the way to describe symmetry properties of the Maxwell equations in  a uniform
media with the use of Lorentz  formulas on the base of  the light velocity  $c$ in
vacuum\footnote{
This point is the most significant   in the context of the above
established the symmetry of the  Maxwell theory in a media under modified
 Lorentz transformations
involving the light velocity in the  media
 $ kc $.}.  In this  work Minkowski
 had  achieved some unification between  Einstein earlier
 analysis and  electrodynamics \underline{in media} in fact.}

Here might be  mentioned specially that logical construction of
Special relativity by Einstein formally does not depend on the
numerical values of the light velocity -- this might be 300000
km/sec  as well as 3 sm/sec. Essential is only the existence of a
(light) signal which goes through the space with the same velocity
for all inertial   observers. And in this context we must
recognize that operating with a light signal of velocity $c$ in a
media is only a  mental fiction; in fact any real  light can move
through a uniform media with velocity $\tilde{c}= kc$. So it might
seem well-taken the requirement to perform clock synchronization
in the  uniform  medias with the help of  real light signals
influenced by the  media. However  this was not done by Einstein
and also  it was not done by  Minkowski. On the contrary,
Minkowski  found the way to speak about relativistic symmetry of
the Maxwell's theory in a media and to use only the Lorentz
formulas with  the vacuum light velocity $c$. Below we will
introduce the Minkowski's   approach  [6]  without following  it
in detail (see also [16,17,18]).

\section{ Standard Lorentz symmetry of the  Maxwell theory
in a media  }

\hspace{5mm} Let us start from the  Maxwell's
equations in the form
 ($x^{0} = ct, J^{0}=\rho $ )
\begin{eqnarray}
\mbox{div} \; {\bf D} =   J^{0} \; ,\qquad
\mbox{rot}\; {{\bf H} \over c}   =  { {\bf J} \over c} \; +\;
{\partial {\bf D} \over \partial  ct} \;     \; , \qquad
\mbox{div} \;  c{ \bf B} = 0 \; , \qquad
\mbox{rot}\; {\bf E}  = - \;
{\partial c{\bf B} \over \partial  ct} \;   \; .
\label{5.2.1}
\end{eqnarray}

\noindent
Here  equations are divided into two  groups:
for vectors $ ({\bf D}, {\bf H} / c)$ and for vectors  $({\bf E}, {c\bf B} )$.
 Note that the source fields
$( J^{0}= \rho,  {\bf J} /c )$ enter only the first group. Also one point to emphasize is
that  eqs. (\ref{5.2.1})  do not include parameters of dielectric and magnetic
penetrability, however as a peculiar compensation for this we need to use concurrently
 two sets of electromagnetic
vectors:  $ ({\bf D}, {\bf H} / c ) $  and  $({\bf E}, {c\bf B} )$.

It is  readily established that if  eqs.  (\ref{5.2.1}) are subjected to
the  (ordinary)  Lorentz transformation
(with the light velocity $c$ in the vacuum  and correspondingly with the
  variable  $x^{0} =ct$):
  \begin{eqnarray}
x^{'0} = \mbox{ch}  \beta   x^{0} - \mbox{sh}\beta  x^{1} \; ,\qquad
x^{'1} = - \mbox{sh}  \beta   x^{0} + \mbox{ch}\beta  x^{1} \; ,\qquad
x^{'2} = x^{2}  , \qquad  x^{'3} = x^{3}\; ,
\nonumber
\\
J^{'0} = \mbox{ch}  \beta   J^{0} - \mbox{sh}\beta  c^{-1} J^{1}  , \qquad
c^{-1} J^{'1} = -\mbox{sh}  \beta   J^{0} + \mbox{ch}\beta  c^{-1} J^{1}  ,\qquad
J^{'2} = J^{2} , \;\;  J^{'3} = J^{3} ,
\nonumber
\\
D^{'1} = + D^{1}  ,    \qquad
D^{'2}= \mbox{ch}  \beta   D^{2} - \mbox{sh}  \beta   c^{-1} H^{3}  ,\qquad
c^{-1} H^{'3}=    - \mbox{sh}  \beta   D^{2} +\mbox{ch}  \beta   c^{-1}H^{3}  ,
\nonumber
\\
H^{'1} = + H^{1}  ,   \qquad
D^{'3}= \mbox{ch}  \beta   D^{3} + \mbox{sh}  \beta   c^{-1} H^{2} , \qquad
c^{-1} H^{\;'2}=    + \mbox{sh}  \beta   D^{3} + \mbox{ch}
 \beta   c^{-1} H^{2}   ,
 \nonumber
 \\
 E^{'1} = + E^{1}  , \qquad
 E^{'2}= \mbox{ch}  \beta   E^{2} - \mbox{sh}  \beta   c B^{3}  , \qquad
c B^{'3}=    -\mbox{sh}  \beta   E^{2} +\mbox{ch}  \beta   c B^{3}  \; ,
\nonumber
\\
B^{'1} = + B^{1} \; ,   \qquad
E^{'3}= \mbox{ch}  \beta   E^{3} + \mbox{sh}  \beta   c B^{2} \; , \qquad
c B^{'2}=    +\mbox{sh}  \beta  \; E^{3} + \mbox{ch}  \beta   c B^{2}  \; ,
\label{5.2.4}
\end{eqnarray}

\noindent we again will obtain equations in the Maxwell's form:
\begin{eqnarray}
\mbox{div} '\; {\bf D}' =   J^{'0} \; ,\qquad
\mbox{rot}'\; {{\bf H}' \over c}   =  { {\bf J}' \over c}  +
{\partial {\bf D}' \over \partial  ct'} \;     , \qquad
\mbox{div} '\;  c{ \bf B}' = 0 \; ,  \qquad
\mbox{rot}'\; {\bf E}'  = - \;
{\partial  c{\bf B}' \over \partial  ct'} \;  .
\label{5.2.5}
\end{eqnarray}

\begin{quotation}

The point of  first importance is that
the modified  Lorentz transformations used in (\ref{5.1.15}) generate much different formulas:
from formal view point all  the
difference is reduced to appearance of the  modified  quantity  $kc$ everywhere instead of $c$:

\end{quotation}

So,  with Maxwell's equation we have  faced a rather peculiar situation
when at the same time  two  different symmetries are revealed:

\noindent
I)  a  symmetry  with respect to  the ordinary Lorentz transformations
  in which there is  presented a  universal constant -- the light velocity in the
 vacuum;

\noindent
II) another symmetry with respect to the modified Lorentz transformations
 in which there is presented
a media dependent constant -- the light velocity in  the media;

\noindent
III) explicit  transforms both  for
space-time coordinates and for electromagnetic quantities
 differ for these two cases.

Which symmetry of   these two  is more  correct or adequate?
What attitude should we have to
the fact itself of existence of two symmetries for Maxwell equations in  a media?
Which of them corresponds more closely to the physical reality?
Exist there any criteria to pick out only one of two logical possibilities?
All these questions should be  answered.

From purely theoretical view point, taking seriously the need to synchronize   clocks
with the help of real light signals  in a media, one must use  the modified
version of the Lorentz transformations in the media (also see [13]).

\section{Field restrain conditions  and  the Minkowski equations }

\hspace{5mm}
Now let us  examine  the following problem:
what form will  the field relations
\begin{eqnarray}
  D^{i} = \epsilon_{0} \epsilon \;  E^{i}, \qquad
H^{i} = {1 \over  \mu_{0} \mu  } \;  B^{i} \;,
\label{5.2.10}
\end{eqnarray}

\noindent take after the Lorentz  transformation to a moving reference  frame?
Firstly, this problem was considered and solved by H. Minkowski in 1908  [6].
For simplicity we will take the most simple Lorentz formulas that correspond to a moving
reference frame along axis $x$.

 In the first place consider the ordinary Lorentz transforms. With the use of (\ref{5.2.4})
 from  eqs. (\ref{5.2.10}) it follows
\begin{eqnarray}
\underline{   D^{i} = \epsilon_{0} \epsilon \;  E^{i}}\; , \qquad \Longrightarrow
D^{'1} = \epsilon_{0} \epsilon   E^{'1}  ,
\nonumber
\\
 \mbox{ch}  \beta   D^{'2} + \mbox{sh}  \beta    {H^{'3}  \over c}  =
 \epsilon_{0} \epsilon  (
\mbox{ch}  \beta   E^{' 2} + \mbox{sh}  \beta   c B^{'3} ) ,
\nonumber
\\
\mbox{ch}  \beta   D^{'3} - \mbox{sh}  \beta   { H^{'2} \over c}=
 \epsilon_{0} \epsilon  (
\mbox{ch}  \beta   E^{;'3} - \mbox{sh}  \beta   c B^{'2} )  ;
\label{5.2.12a}
\\
\underline{H^{i} = {1 \over  \mu_{0} \mu  } \;  B^{i}}\; , \qquad \Longrightarrow
\qquad
H^{'1} = {1 \over  \mu_{0} \mu  }   B^{1}  ,
\nonumber
\\
\mbox{sh}  \beta   D^{'3} - \mbox{ch}  \beta   {H^{'2} \over  c}  =
{1 \over  \mu_{0} \mu  }  {1 \over  c^{2} }\;(
\mbox{sh}  \beta   E^{'3}   - \mbox{ch}  \beta    c B^{'2}  ) ,
\nonumber
\\
\mbox{sh}  \beta    D^{'2} +\mbox{ch}  \beta   {H^{'3}\over  c}  =
{1 \over  \mu_{0} \mu }{1\over   c^{2} }\; (\mbox{sh}  \beta
E^{'2} + \mbox{ch}  \beta   c B^{'3})  .
 \label{5.2.12b}
\end{eqnarray}

\noindent Relations  (\ref{5.2.12a}) and (\ref{5.2.12b})  are just
what we  call \underline{Minkowski equations}   [6]  written down
in a particular simple case. Let us change  them  to another form.
To this end, they should be  rewritten as three pairs of linear
systems under the variables
 $(D^{\;'1},H^{\;'1}),
(D^{\;'2},H^{\;'3}/c), (D^{\;'3},H^{\;'2}/c)$.
Solutions of   which look as follows
\begin{eqnarray}
D^{'1} = \epsilon_{0} \epsilon \;   E^{'1}  \;,
\nonumber
\\
D^{'2}=
 \epsilon_{0} \epsilon \;  [ \; (\mbox{ch}^{2} \beta - k^{2}  \mbox{sh}^{2} \beta )\;
E^{'2} +
 \mbox{sh} \beta  \;  \mbox{ch} \beta   \; (1 -k^{2} )  \; c B^{'3}
 \; ] \;  ,
\nonumber
\\
D^{'3}=
\epsilon_{0} \epsilon  \;  [\;  (\mbox{ch}^{2} \beta - k^{2}  \mbox{sh}^{2} \beta )\;
E^{'3} -
 \mbox{sh} \beta  \; \mbox{ch} \beta  \;  (1 -k^{2} ) \;  c B^{'2} \;
]\;   ,
\nonumber
\\
H^{'1}/  c  =  \epsilon_{0} \epsilon k^{2}   \;  c B^{1}  \; ,
\nonumber
\\
H^{'2} /  c  =
\epsilon_{0} \epsilon  \;  [ \; (k^{2} \mbox{ch}^{2} \beta - \mbox{sh}^{2} \beta )\;
 cB^{'2} -
  \mbox{sh} \beta \;  \mbox{ch} \beta \;   (k^{2} -1 ) \;  E^{'3}\;   ],
 \nonumber
  \\
H^{'3} /c   =
\epsilon_{0} \epsilon \;  [\;   (k^{2}\;  \mbox{ch}^{2} \beta - \mbox{sh}^{2} \beta )\;
 cB^{\;'3} +
  \mbox{sh} \beta  \; \mbox{ch} \beta   \; (k^{2} -1 ) \;  E^{'2}\;    ] \;  .
\label{5.2.18}
\end{eqnarray}

\noindent Equations  (\ref{5.2.18}) say that simple connections (\ref{5.2.10})  between electromagnetic vectors   in initial
(unmoving)  frame  after  translating to a moving  reference frame
 become rather  complex ones -- they  involve now  the velocity as a parameter.
In other terms, this means that the field relations  (\ref{5.2.10})
are not Lorentz invariant.
However,  we  can see that in the vacuum case when
 $k=1$ the  formulas  (\ref{5.2.18})  will take the same  most simple form from which we have started
 initially:
\begin{eqnarray}
\underline{k=1}\;,    \;  \;
D^{\;'i} = \epsilon_{0}  \;  E^{\;'i} \; , \;
H^{\;'i} = {1 \over  \mu_{0}  } \;  B^{\;i} \; .
\label{5.2.19}
\end{eqnarray}

Now, let us  proceed to the consideration of the  the field  relations when we use
the modified Lorentz  transformations. We will easily see that  properties of these
 relations
under that modified Lorentz theory  become different and much more  attractive: they  turn out to be
Lorentz invariant\footnote{This is a long time known result, see for instance the
 Rosen's work  [13]}.
Indeed, let us  start with (everywhere instead of  $c$  there
appears  $kc$ ):
\begin{eqnarray}
\underline{   D^{i} = \epsilon_{0} \epsilon \;  E^{i} }\; , \qquad \Longrightarrow \qquad
D^{\;'1} = \epsilon_{0} \epsilon \;  E^{\;'1} \; ,
\nonumber
\\
\mbox{ch}  \sigma  \; D^{\;'2} + \mbox{sh}  \sigma  \;  {H^{\;'3}  \over kc}  =
 \epsilon_{0} \epsilon \; (
\mbox{ch}  \sigma  \; E^{\;' 2} + \mbox{sh}  \sigma  \; kc B^{\;'3} )\; ,
\nonumber
\\
\mbox{ch}  \sigma  \; D^{\;'3} - \mbox{sh}  \sigma  \; { H^{\;'2} \over kc}=
\epsilon_{0} \epsilon \; (
\mbox{ch}  \sigma  \; E^{\;'3} - \mbox{sh}  \sigma  \; kc B^{\;'2} ) \; ,
\label{5.2.20a}
\\
\underline{H^{i} = {1 \over  \mu_{0} \mu  } \;  B^{i}}
\; , \qquad \Longrightarrow \qquad
H^{\;'1} = {1 \over  \mu_{0} \mu  } \;  B^{\;'1} \; ,
\nonumber
\\
-\mbox{sh}  \sigma  \; D^{\;'3} + \mbox{ch}  \sigma  \; {H^{\;'2} \over k c}  =
{1 \over  \mu_{0} \mu  } \; {1 \over  k^{2}c^{2} }\;(
-\mbox{sh}  \sigma  \; E^{\;'3}   + \mbox{ch}  \sigma  \;  kc B^{\;'2}  )\; ,
\nonumber
\\
\mbox{sh}  \sigma  \;  D^{\;'2} +\mbox{ch}  \sigma  \; {H^{\;'3}\over k c}  =
{1 \over  \mu_{0} \mu } \; {1 \over k^{2}c^{2} }\;(
\mbox{sh}  \sigma  \; E^{\;'2} + \mbox{ch}  \sigma  \; kc B^{\;'3}) \; .
\label{5.2.20b}
\end{eqnarray}

\noindent Their solutions are
\begin{eqnarray}
D^{\;'i}=
 \epsilon_{0} \epsilon \;  E^{\;'i} \; ,\qquad
H^{\;'i} = {1 \over \mu_{0} \mu}\;  B^{\;'i} \;.
\label{5.2.26}
\end{eqnarray}

So we have arrived at the most attractive result from theoretical viepoint:
The Maxwell's in  the media together with  the field restrain conditions
  turn out to be invariant under {\em modified} \hspace{2mm}
 Lorentz  transformations. This is a significant theoretical argument in favor of
the Lorentz   symmetry involving the  light velocity in a media $kc$ instead of the light velocity
in the  vacuum $c$.
Such a modified theoretical scheme looks  simpler and more attractive then commonly
exploited one.

\section{4-tensor formalism}

\hspace{5mm}
The two  Maxwell equations with sources
\begin{eqnarray}
 \mbox{div} \; {\bf D} =   J^{0} \; ,\qquad
 \mbox{rot}\; {{\bf H} \over c}   =  { {\bf J} \over c} \; +\;
  {\partial {\bf D} \over \partial  ct} \;
\nonumber
\end{eqnarray}

\noindent can be  presented in a very compact and simple  form
if  one introduces a special notation with the  use  of indices taking over four values.
Let us introduce the tensor
\begin{eqnarray}
(H^{ab}) = \left | \begin{array}{cccc}
0  & -D^{1}  & -D^{2} & -D^{3} \\
+D^{1}   &  0  &  -H^{3}/c  & +  H^{2}/c \\
+D^{1} & + H^{3}/c  & 0 & - H^{1}/c \\
+D^{3} & - H^{2}/c & +H^{1}/c & 0
\end{array} \right | .
\label{5.2.29}
\end{eqnarray}

\noindent
The main  assertion is that the above two  equations  are  equivalent  to the  tensor one
(the notation
$x^{a} = (ct; \; x^{i}), \; j^{\;a} = (J^{0} ,  J^{i} / c )$
is used)
\begin{eqnarray}
\partial _{b}  \; H^{ba}    =  j^{a}\; .
\label{5.2.31}
\end{eqnarray}

Now let us consider the two remaining Maxwell equations
\begin{eqnarray}
\mbox{div} \;  c{ \bf B} = 0 \; , \qquad
 \mbox{rot}\; {\bf E}  = - \;
{\partial  c{\bf B} \over \partial  ct} \;   \; .
\nonumber
\end{eqnarray}

\noindent
To deal with these two equations Minkowski had introduced  [6] another  tensor $F^{ab}$:
\begin{eqnarray}
(F^{ab}) = \left | \begin{array}{cccc}
0  & -E^{1}  & -E^{2} & -E^{3} \\
+E^{1}   &  0  &  -cB^{3}  & +  cB^{2}\\
+E^{1} & + cB^{3}  & 0 & - cB^{1} \\
+E^{3} & - cB^{2}& + cB^{1}& 0
\end{array} \right | \; .
\label{5.2.33}
\end{eqnarray}

\noindent
The main assertion here is that the  remaining Maxwell's equations
are equivalent to the  tensor one
\begin{eqnarray}
\partial_{c}\;  F_{ab} +  \partial_{a}\;  F_{bc} + \partial_{b}\;  F_{ca}  = 0 \; .
\label{5.2.34}
\end{eqnarray}

Thus, we  have arrived at the  compact tensor form of the  Maxwell equations:
\begin{eqnarray}
\partial _{b}  \; H^{ba}    =  j^{a}\; , \qquad
\partial_{c}\;  F_{ab} +  \partial_{a}\;  F_{bc} + \partial_{b}\;  F_{ca}  = 0 \; .
\label{5.2.35}
\end{eqnarray}

The Maxwell's equations in other variables, in which they exhibit
symmetry under modified Lorentz  transformations,
 may be  rewritten with the use of only one
electromagnetic tensor $(f^{AB}) = ({\bf d} , {\bf h}) $ (
appearance here and in the following of
 the  capital letters  to stand for tensor's indexes means that such quantities  transform
 in accordance with the modified Lorentz  symmetry)
in the form of two tensor equations
\begin{eqnarray}
\partial _{B}  \; f^{BC}    =  j^{C}\; , \qquad
\partial_{C}\;  f_{AB} +  \partial_{A}\;  f_{BC} + \partial_{B}\;  f_{CA}  = 0 \; .
\label{5.2.36c}
\end{eqnarray}

\noindent Besides, the  Maxwell's equations, invariant under
modified Lorentz transformations, may be rewritten with the help
of two tensors as well. Indeed,  equations (\ref{5.1.9}) may be
taken as
\begin{eqnarray}
\mbox{div} \;  kc{ \bf B} = 0 \; , \;
 \mbox{rot}\; {\bf E}  = - \;
{\partial \over \partial  kct} \; kc{\bf B}  \; , \qquad
 \mbox{div} \; {\bf D} =   J^{0} \; ,\;
 \mbox{rot}\; {{\bf H} \over kc}   =  { {\bf J} \over kc} \; +\;
  {\partial \over \partial  kct} \;  {\bf D}   \; .
\label{5.2.37a}
\end{eqnarray}

\noindent
From where, introducing (modified) electromagnetic tensors:
\begin{eqnarray}
(H^{AB}) = ( {\bf D}, {\bf H} / kc ) \; , \;
(F^{AB}) = ( {\bf E}, kc {\bf B}  )
\label{5.2.37b}
\end{eqnarray}

\noindent  eqs. (\ref{5.2.37a})   can be  readily presented as follows:
\begin{eqnarray}
\partial _{B}  \; H^{BA}    =  j^{A}\; , \qquad
\partial_{C}\;  F_{AB} +  \partial_{A}\;  F_{BC} + \partial_{B}\;  F_{CA}  = 0 \; .
\label{5.2.37c}
\end{eqnarray}

\section{Minkowski  relations in covariant  tensor form}

\hspace{5mm}
 The Minkowski's equations
 may be  quite easily rewritten in a special  form that remains the same for any  Lorentz transformations, including  arbitrary rotations and (uniform) movings.
 Trick  that is  employed below   is simple but  useful  and often applicable.
It is based on the following property of the tensor formalism:
 {\em if we think (know)  that a certain physical  equation   must be Lorentz invariant
 and  an explicit form of the  equation is given only in some particular  reference frame
 then its invariant  form  may be  found   with the help of Lorentz  transformations.
   The same may  be achieved   if we can see in the particular  equation its general tensor
 form.
}

To make use of this trick, one special notion, 4-vector of velocity  that can be related
to the moving media,  is needed
\begin{eqnarray}
u^{a} = {dx^{a} \over ds } =  ( {1 \over \sqrt{1 - v^{2} / c^{2}} }, {v^{i}/c \over
\sqrt{1 - v^{2} / c^{2}} }) \; .
\label{5.2.52b}
\end{eqnarray}

\noindent
To obtain a  tensor form of Minkowski equations, we will take a particular 4-velocity vector:
\begin{eqnarray}
u^{a} =
( {1 \over \sqrt{1 - v^{2} / c^{2}} }, \;{- v/c \over
\sqrt{1 - v^{2} / c^{2}} }, \; 0, \; 0) \;  =
 ( \mbox{ch}\beta, - \mbox{sh}\beta,0,0)\; . \qquad \qquad
\label{5.2.53}
\end{eqnarray}

\noindent
It is the matter of simple calculation to see [6] that
 the all six Minkowski  equations
are equivalent to the tensor ones:
\begin{eqnarray}
H^{ab} u_{b} = \epsilon_{0}\epsilon F^{ab} u_{b} \;, \qquad \qquad \qquad
\label{5.2.55a}
\\
H^{ab} u^{c} + H^{bc} u^{a} + H^{ca} u^{b} =
  {1  \over  c^{2} \mu \mu_{0} } \;
(F^{ab} u^{c} + F^{bc} u^{a} + F^{ca} u^{b})  \; .
\label{5.2.55b}
\end{eqnarray}

In the  vacuum  case, when $\epsilon=1, \mu =1$,
eqs. (\ref{5.2.55a})  and  (\ref{5.2.55b}) may be rewritten differently
\begin{eqnarray}
H^{ab} u_{b} = \epsilon_{0} F^{ab} u_{b} \;, \qquad \qquad \qquad
\label{5.2.56a}
\\
H^{ab} u^{c} + H^{bc} u^{a} + H^{ca} u^{b} =  \epsilon_{0}
(F^{ab} u^{c} + F^{bc} u^{a} + F^{ca} u^{b} ) \; .
\label{2.5.56b}
\end{eqnarray}

\noindent
These  tensor equations  admit a simple solution. Indeed,
let us  multiply eq. (\ref{2.5.56b}) by $u_{c}$ (considering  $u^{c}u_{c} =+1$), then
\begin{eqnarray}
H^{ab} = \epsilon_{0}
(F^{ab}  + F^{bc} u_{c} u^{a} + F^{ca} u_{c} u^{b} ) -
 H^{bc} u_{c} u^{a} - H^{ca} u_{c} u^{b} \; ;
\nonumber
\end{eqnarray}

\noindent
from where, bearing in mind  (\ref{5.2.56a}),  we arrive at
\begin{eqnarray}
H^{ab} =
\epsilon_{0} F^{ab}\; .
\label{5.2.57}
\end{eqnarray}

\noindent
This tensor condition, in component form will  look as six relations
(just  the same  ones were obtained earlier in    (\ref{5.2.19}))
\begin{eqnarray}
D^{i}  = \epsilon_{0} \; E^{i}\; , \qquad
H^{i} = {1 \over  \mu_{0}  } \; B^{i} \; .
\nonumber
\end{eqnarray}

However, for any  media, analogous  calculation leads
to a very different result. Indeed, let us multiply
(\ref{5.2.55b}) by   $u_{c}$:
\begin{eqnarray}
H^{ab} + H^{bc} u_{c}  u^{a} + H^{ca} u_{c}  u^{b} =
 {1 \over  c^{2} \mu \mu_{0} } \; (
F^{ab}  + F^{bc} u_{c}  u^{a} + F^{ca} u_{c}  u^{b} )  \; .
\nonumber
\end{eqnarray}

\noindent  From this, taking  (\ref{5.2.55a}), we get to\begin{eqnarray}
H^{ab}  =  \epsilon_{0}\epsilon k^{2}  \;
F^{ab}  +
 \epsilon_{0}\epsilon  \; (k^{2} -1) \;
 [ \;   F^{bc} u_{c}  u^{a} -  F^{ac} u_{c}  u^{b} \; ]   \; .
\label{5.2.58}
\end{eqnarray}

\noindent
Evidently, these are a covariant tensor  form the  earlier found   in (\ref{5.2.18}) --
also see in [7-10,13,14,16-18].

Take notice that while using the Maxwell theory with only one (modified) tensor
 $f^{AB}$,
  no  additional condition between electromagnetic tensors is needed at all.
Some clarifying  analysis  can be easily done.
In this case, there arise modified Minkowski relations ( $c$  is changed by
$kc$ ):
\begin{eqnarray}
D^{i} = \epsilon_{0} \epsilon \;  E^{i}  :   \qquad \Longrightarrow \qquad
D^{\;'1} = \epsilon_{0} \epsilon \;  E^{\;'1} \; ,
\nonumber
\\
\mbox{ch}  \sigma  \; D^{\;'2} + \mbox{sh}  \sigma  \;  {H^{\;'3}  \over kc}  =
 \epsilon_{0} \epsilon \; (
\mbox{ch}  \sigma  \; E^{\;' 2} + \mbox{sh}  \sigma  \; kc B^{\;'3} )\; ,
\nonumber
\\
\mbox{ch}  \sigma  \; D^{\;'3} - \mbox{sh}  \sigma  \; { H^{\;'2} \over kc}=
 \epsilon_{0} \epsilon \; (
\mbox{ch}  \sigma  \; E^{\;'3} - \mbox{sh}  \sigma  \; kc B^{\;'2} ) \; ,
\label{5.2.60a}
\\
H^{i} = {1 \over  \mu_{0} \mu  } \;  B^{i}: \; \Longrightarrow \qquad
H^{\;'1} = {1 \over  \mu_{0} \mu  } \;  B^{\;1} \; ,
\nonumber
\\
-\mbox{sh}  \sigma  \; D^{\;'3} + \mbox{ch}  \sigma  \; { H^{\;'2} \over k c}  =
{1 \over  \sigma_{0} \mu  } \; {1 \over  k^{2}c^{2} }\;(
-\mbox{sh}  \sigma  \; E^{\;'3}   + \mbox{ch}  \sigma  \;  kc B^{\;'2}  )\; ,
\nonumber
\\
\mbox{sh}  \sigma  \;  D^{\;'2} +\mbox{ch}  \sigma  \; {H^{\;'3}\over k c}  =
 {1 \over  \mu_{0} \mu } \; {1 \over k^{2}c^{2} }\;(
\mbox{sh}  \sigma  \; E^{\;'2} + \mbox{ch}  \sigma  \; kc B^{\;'3}) \; .
\label{5.2.60b}
\end{eqnarray}

\noindent
Now a modified 4-velocity $U^{A}$   is needed:
\begin{eqnarray}
U^{A} =
 {dx^{a} \over ds } =
  ( {1 \over \sqrt{1 - v^{2} / k^{2}c^{2}} }, {v^{i}/kc \over
\sqrt{1 - v^{2} / k^{2}c^{2}} }) \;,
\label{5.2.61a}
\end{eqnarray}

\noindent
and its particular form
\begin{eqnarray}
U^{A} =
( {1 \over \sqrt{1 - v^{2} / k^{2} c^{2}} }, {- v/kc \over
\sqrt{1 - v^{2} / k^{2} c^{2}} }, \; 0, \; 0) \;  =
  ( \mbox{ch}\sigma, - \mbox{sh}\sigma,0,0)\; .
\nonumber
\end{eqnarray}

\noindent
Tensor representation of eqs.  (\ref{5.2.60a}) and   (\ref{5.2.60b}) is
\begin{eqnarray}
H^{AB} U_{B} = \epsilon_{0}\epsilon F^{AB} U_{B} \; ,
\label{5.2.62a}
\end{eqnarray}
\begin{eqnarray}
H^{AB} U^{C} + H^{BC} U^{A} + H^{CA} U^{B} =
 {1 \over  k^{2}  c^{2} \mu \mu_{0} }
(F^{AB} U^{C} + F^{BC} U^{A} + F^{CA} U^{B} )  \; .
\label{5.2.62b}
\end{eqnarray}

\noindent The latter equation may be rewritten as
\begin{eqnarray}
H^{AB} U^{C} + H^{BC} U^{A} + H^{CA} U^{B} =
 \epsilon_ {0} \epsilon \;
(F^{AB} U^{C} + F^{BC} U^{A} + F^{CA} U^{B} )  \; .
\label{5.2.62c}
\end{eqnarray}

\noindent
Multiplying it by  $U_{C}$:
\begin{eqnarray}
H^{AB}  + H^{BC} U_{C} U^{A} + H^{CA}  U_{C} U^{B} =
 \epsilon_ {0} \epsilon \;
(F^{AB}  + F^{BC} U_{C}  U^{A} + F^{CA}  U_{C} U^{B} )  \; ,
\nonumber
\end{eqnarray}

\noindent
from where, with  (\ref{5.2.62a}),  it follows
\begin{eqnarray}
H^{AB}   = \epsilon_ {0} \epsilon \; F^{AB} \; , \qquad
\mbox{or}\qquad D^{i}  = \epsilon_{0}\epsilon  \; E^{i} \; ,
 \qquad
H^{i} = {1 \over  \mu_{0} \mu } \; B^{i} \; . \label{5.2.63a}
\end{eqnarray}

\section{Potentials in a media }

\hspace{5mm}
In remains to be seen which peculiarities arise from the  presence of a
 uniform media  when we try do describe electromagnetic fields in terms
 of the  scalar and vector potentials $\varphi, {\bf A}$:
 $$
\left \{ \; {\bf E}  , \; {\bf B} , \;
{\bf D} , \; {\bf H} \; \right \} \;
\Longrightarrow \qquad
\left \{\;  \varphi  , \;  {\bf A}  \; \right \} \; .
$$

We might anticipate some difficulties because  the
Maxwell theory  in the media  exhibits  symmetry  under ordinary,
the vacuum light velocity based, Lorentz  transformations only
if the  two  electromagnetic tensors, $H^{ab}$ and $F^{ab}$, are used.
There is no ground for feeling enthusiastic about existence of relativistically invariant
 formulas (\ref{5.2.55a}) and  (\ref{5.2.55b}).
 These formulas  involve 4-velocity vector, characteristic of the state
 of moving of the  media under the  inertial reference frame. In other words,
 these formulas may be   understood  as explicit dependence of the basic electrodynamic
  equations  upon the absolute velocity of  a moving body.
It somehow contradicts to the initial principles and main claims of Special relativity theory.

Let us write down the Maxwell's equations again:
\begin{eqnarray}
\nabla \bullet \;  {\bf B} = 0  ,
\qquad
\nabla \times \; {\bf E}  = -  {\partial  {\bf B}\over
\partial t}  ,
\nonumber
\\
\nabla \bullet \; {\bf E} =  {1 \over  \epsilon \epsilon_{0}}  \rho  , \qquad
{1 \over  \mu \mu_{0}  }    \nabla \times\; {\bf B}  =   {\bf J}
+\; \epsilon \epsilon_{0}   {\partial   {\bf E} \over \partial t}  .
\label{5.2.78b}
\end{eqnarray}

\noindent
The  most general substitution for potentials   $ \varphi ,\;
 {\bf A} \;$
 which makes two first equations  (\ref{5.2.78b}) identities has the form
 \begin{eqnarray}
{\bf B} = d  \;   \nabla \times  {\bf A} , \;\;
{\bf E}  = - n \; \nabla \varphi - d \; {\partial  {\bf A}  \over  \partial t} \;\; ,
\label{5.2.79}
\end{eqnarray}

\noindent where  $d,n$ are some yet unknown parameters. The  first
equation with the source $\rho$ in   (\ref{5.2.78b}) gives
\begin{eqnarray}
(\; -\; \nabla^{2} \varphi +  {dm\over n}  \; {\partial^{2} \varphi \over
\partial t^{2} }  \;  )  =
{1 \over  n \epsilon \epsilon_{0}} \; \rho
+
{d \over n} \; {\partial \over  \partial t} \;  ( \; \nabla \bullet
{\bf A} + m {\partial \varphi  \over \partial t } \;   )\; .
\label{5.2.80a}
\end{eqnarray}

\noindent
The  second equations in  (\ref{5.2.78b})  with the source ${\bf J}$ leads to
\begin{eqnarray}
 \nabla \times \;    ( \nabla \times  {\bf A})
  =  {\mu \mu _{0}\over d} \; {\bf J}
+  {\epsilon \epsilon_{0} \mu \mu_{0} \over d }  \;  {\partial \over \partial t} \;
(- n\;\nabla \varphi - d \; {\partial  {\bf A} \over  \partial t} \;  ) \; .
\nonumber
\end{eqnarray}

\noindent From this, with the help of the identity
$
 \nabla\times (\nabla \times {\bf A}) =
 - \nabla^{2}  \; {\bf A}  + \nabla \; ( \nabla \bullet {\bf A})  \;,
$
we get to
\begin{eqnarray}
 ( \; - \nabla^{2}  \; {\bf A}  + \epsilon_{0}\epsilon \mu_{0}\mu
\; {\partial ^{2} \over  \partial t^{2}} \; {\bf A} \;  ) =
{\mu \mu_{0} \over d} \;
{ \bf J} - \nabla \;   ( \;  \nabla \bullet {\bf A}
 +
{\epsilon_{0}\epsilon \mu_{0}\mu \; n \over d} \;
{\partial  \varphi\over \partial t} \;   ) \; .
\label{5.2.80b}
\end{eqnarray}

\noindent
Comparing  (\ref{5.2.80a}) and  (\ref{5.2.80b}), we see that it suffices to
demand
\begin{eqnarray}
{dm \over n} = \epsilon_{0}\epsilon \mu_{0}\mu   \;, \qquad m =
{\epsilon_{0}\epsilon \mu_{0}\mu \; n \over d}
\label{5.2.81a}
\end{eqnarray}

\noindent so that eqs.  (\ref{5.2.80a}) and  (\ref{5.2.80b}) will have quite symmetrical form with the same
wave operator  on the  left:
\begin{eqnarray}
 (\; -\nabla^{2} \varphi +  \epsilon_{0}\epsilon \mu_{0}\mu \; {\partial^{2} \varphi \over
\partial t^{2} }  \;  )  =
{1 \over  n \; \epsilon \epsilon_{0}} \; \rho \;
+ {d \over n} \;  {\partial \over  \partial t} \;   ( \; \nabla \bullet
{\bf A} + m {\partial \varphi  \over \partial t }  \;  ) \;,
\label{5.2.81b}
\\
 ( \; - \nabla^{2}  \; {\bf A}  + \epsilon_{0}\epsilon \mu_{0}\mu
\; {\partial ^{2}{\bf A}  \over  \partial t^{2}} \;  ) =
 {\mu \mu_{0} \over d} \;
{ \bf J} - \nabla \;   ( \;  \nabla \bullet {\bf A} + m\;
{\partial  \varphi\over \partial t} \;   ) \; .
\label{5.2.81c}
\end{eqnarray}

\noindent With the use of $c$ and $k$, the previous equations
become
\begin{eqnarray}
{1 \over k^{2} c^{2}} = { d m \over n}  \;, \qquad \qquad
\label{5.2.82a}
\\
 (\; -\nabla^{2} \varphi +  {1 \over k^{2} c^{2}} \; {\partial^{2} \varphi \over
\partial t^{2} }  \; )  =
{1 \over  n\; \epsilon \epsilon_{0}} \; \rho \;
+ {d \over n}  \;  {\partial \over  \partial t} \;   ( \; \nabla \bullet
{\bf A} + m {\partial \varphi  \over \partial t }  \;  ) \; ,
\label{5.2.82b}
\\
( \; - \nabla^{2}  \; {\bf A}  + {1 \over k^{2} c^{2}}
\; {\partial ^{2}{\bf A}  \over  \partial t^{2}} \;  ) =
 {\mu \mu_{0} \over d} \;
{ \bf J} - \nabla \;   ( \;  \nabla \bullet {\bf A} + m\;
{\partial  \varphi\over \partial t} \;   ) \; .
\label{5.2.82c}
\end{eqnarray}

\noindent
Relationship  (\ref{5.2.82a}) allows different solutions. The most symmetrical and simplified all
formulas  seems to be
\begin{eqnarray}
m = {1 \over kc} , \; n = {1 \over  \epsilon_{0} \epsilon } \; , \qquad
d = {n \over kc } = {1 \over kc \; \epsilon_{0} \epsilon } \; ,
\qquad
{\mu \mu_{0} \over d} = \mu \mu_{0} \epsilon \epsilon_{0} \;   kc = {1 \over kc } \; . \qquad
\label{5.2.83a}
\end{eqnarray}

\noindent at this eqs.  (\ref{5.2.82b}) and   (\ref{5.2.82c})provide us with these
\begin{eqnarray}
 ( -\nabla^{2} \varphi +  {1 \over k^{2} c^{2}}  {\partial^{2} \varphi \over
\partial t^{2} }   )  =
 \rho
+ {1 \over kc }    {\partial \over  \partial t}    ( \; \nabla \bullet
{\bf A} + {1 \over kc }  {\partial \varphi  \over \partial t }    ) ,
\nonumber
\\
(  - \nabla^{2}   {\bf A}  + {1 \over k^{2} c^{2}}
 {\partial ^{2}{\bf A}  \over  \partial t^{2}}   ) =
{{ \bf J} \over kc} - \nabla    (   \nabla \bullet {\bf A} + {1 \over kc }
{\partial  \varphi\over \partial t}    )  .
\label{5.2.83c}
\end{eqnarray}

\noindent
Eqs.     (\ref{5.2.83c})   may be rewritten as  a (modified)  tensor equation:
\begin{eqnarray}
\partial^{B} \partial_{B} \; A^{C} =   j^{C} +
\partial^{C} (\partial_{B}A^{B}) \; ,
\label{5.2.84}
\end{eqnarray}

\noindent
where
\begin{eqnarray}
 x^{C} = (kct , x^{i}) \; , \; A^{C} =(\varphi, A^{i}),\;
  j^{C} = (\rho, \; {J^{i} \over kc } ) \; .
\end{eqnarray}

\noindent Eq.  (\ref{5.2.84})  proves its invariance under
modified Lorentz transformations construct\-ed on the
base of the light velocity $kc$ in the media. Initial relations  (\ref{5.2.79}) introducing
electromagnetic potentials:
\begin{eqnarray}
\nabla \times  {\bf A} =  {{\bf B} \over \mu_{0} \mu \; kc}  =
{\bf h}   \; , \qquad
-  \nabla \varphi -    {1 \over  kc}\; {\partial  {\bf A} \over
\partial t} = \epsilon_{0} \epsilon  \;  {\bf E}  = {\bf d} \;
\qquad \label{5.2.85a}
\end{eqnarray}

\noindent may be readily translated to tensor form:
\begin{eqnarray}
f_{BC} = \partial_{B} A_{C} - \partial_{C} A_{B} \; ,
\label{5.2.85b}
\end{eqnarray}

\noindent
where  $f_{ab}$ is the electromagnetic tensor for Maxwell equations in modified variables.
Tensor relationship  (\ref{5.2.85b})  permits quite easily reveal a gauge freedom
in determining of electromagnetic potentials:
\begin{eqnarray}
A'_{B} = A_{B} + \partial_{B} \Lambda \; , \qquad \Longrightarrow \qquad
f'_{BC} = f_{BC} \; .
\label{5.2.86}
\end{eqnarray}

Often the Lorentz gauge condition is taken  to be the
most convenient
$
\partial_{B} A^{B} = 0 \; $.
Thus, as a result of the change of variables:
$
{\bf E}, {\bf D},  {\bf B}, {\bf H} \; \Longrightarrow \; (\varphi, {\bf A})=A^{C}
$
the simplicity has been achieved: all the Maxwell's electrodynamics
formally is equivalent to the single equation for 1-rank tensor $A^{B}$.
In the Lorentz gauge, the Maxwell's electrodynamics looks the most simple and  beautiful:
namely it reduces to the  wave  equation:
\begin{eqnarray}
\partial^{B} \partial_{B} \; A^{C} =   j^{C}  \; , \qquad \partial_{B} A^{B}= 0 \; .
\label{5.2.90}
\end{eqnarray}

\noindent  One  notice again:  solutions of the equation (\ref{5.2.90})
there correspond to wave  processes propagating  with the  speed of light
in the media, not in the vacuum, and this velocity is invariant under  modified  Lorentz formulas.

The choice of solution in (\ref{5.2.82a}) that was taken above is not unique. Equally, starting
from (\ref{5.2.82a}),(\ref{5.2.82b}),(\ref{5.2.82c}),
 we may chose the  more traditional one:
\begin{eqnarray}
d = 1 , \qquad n = 1 , \qquad m = {1 \over k^{2} c^{2}} \; ,
\label{5.2.91b}
\\
 (\; -\nabla^{2} \varphi +  {1 \over k^{2} c^{2}} \; {\partial^{2} \varphi \over
\partial t^{2} }  \; )  =
 {1 \over   \epsilon \epsilon_{0}} \; \rho \;
+   {\partial \over  \partial t} \;   ( \; \nabla \bullet
{\bf A} + {1 \over k^{2} c^{2}} \; {\partial \varphi  \over \partial t }  \;  ) ,
\nonumber
\\
( \; - \nabla^{2}  \; {\bf A}  + {1 \over k^{2} c^{2}}
\; {\partial ^{2}{\bf A}  \over  \partial t^{2}} \;  ) =
 \mu \mu_{0}  \;
{ \bf J} - \nabla \;   ( \;  \nabla \bullet {\bf A} + {1 \over k^{2} c^{2}} \;
{\partial  \varphi\over \partial t} \;   )
\nonumber
\end{eqnarray}

\noindent which coincides with that used in the
 the known handbook on electrodynamics by Stratton [1].
 Evidently,
two variants are totally equivalent, because they differ only in determining units for
potentials -- see (\ref{5.2.79}).

\section{Potentials in a media,   ordinary Lorentz symmetry treatment
}

\hspace{5mm}
Now we consider an alternative way of introducing
potentials into electrodynamics in presence of  a media which has its origin in earlier
investigation by Minkowski [6] (just this variant
is being used mainly; see for instance the  thorough review [16].)

 The most noticeable feature of this method consists in
 the following: in this approach we are able to formulate the Maxwell's electrodynamics
 in a media in terms of potentials with the use of the  ordinary
  Lorentz symmetry based on the vacuum light velocity $c$
 only. Concurrently, the mathematical equations achieved look more
 complicated, and also  these equations involve explicitly the velocity of the media
 under the reference frame.
The latter might be considered as return to prehistory of Special relativity
with all searching some absolute velocities.
However, we  are not going to  be  submerged  in so metaphysical subtleties.
Nevertheless,   one point should be emphasized: {\em
there exist two alternative ways to  develop potential approach for electrodynamics in
media -- one developed in previous Section, and another exposed below.
The ways are completely equivalent in mathematical sense.
The first is much more simpler technically  but it presumes invariance
under modified Lorentz symmetry based on the light velocity $kc$.
Why  must we employ the more compli\-cated  technique  -- only because of
its concomitant  ordinary Lorentz symmetry treatment?}

Let us  write  down the Minkowski's  equations
are equivalent to
\begin{eqnarray}
H_{ab} = \Delta_{abmn} \; F^{mn} \; , \qquad
\Delta_{abmn} =
\epsilon_{0}\epsilon k^{2} \; g_{am}\;  g_{bn} +
 \epsilon_{0}\epsilon (k^{2}-1) \;
u_{n} (g_{bm} u_{a} - g_{am} u_{b} ) \;.
\label{5.2.94b}
\end{eqnarray}

Firstly,  the 4-rank tensor connecting   $H^{ab}$  and   $F^{ab}$ was introduced
(for a more general case of anisotropic  media) by Tamm and Mandel'stam [7,8].
 For a  uniform media,  accordingly Watson-Yauch-Riazanov [9,10]
  the tensor $\Delta_{abmn}$ (\ref{5.2.94b})   may be taken in another form:
\begin{eqnarray}
H_{ab} = \Delta_{abmn} \; F^{mn} \; , \qquad
\Delta_{abmn} = A \; (g_{am} + B u_{a}u_{m} ) \; (g_{bn} + B u_{b}u_{n}) \; .
\label{5.2.95b}
\end{eqnarray}

\noindent Although, this  $\Delta_{abmn}$  contains a term of fourth order in
velocity, only terms of second order  give non-zero contribution into the formula
(\ref{5.2.94b}).
Let us demonstrate that   (\ref{5.2.94b})   and (\ref{5.2.95b})   are the same at specially given  $A$ and $B$.
From  (\ref{5.2.95b}) it follows
\begin{eqnarray}
H_{ab} =  A \; g_{am}  \; g_{bn} \;  F^{mn} + AB \; g_{am} B u_{b}u_{n} \; F^{mn} +
\nonumber
\\
+
AB \; u_{a}u_{m} g_{bn}  \; F^{mn} =
 A \; g_{am}  \; g_{bn} \;  F^{mn} +
  AB\;  u_{n}  \; ( g_{am} u_{b} -
g_{bm} u_{a}) \;F^{mn} \; .
\nonumber
\end{eqnarray}

\noindent
Comparing  this with  (\ref{5.2.94b}),  we get
$
A = \epsilon_{0}\epsilon k^{2} \; , \;\;   -AB =
\epsilon_{0}\epsilon (k^{2} -1) \;,
$
from where it follows
\begin{eqnarray}
A = \epsilon_{0}\epsilon k^{2}  , \qquad  B=  {1 - k^{2} \over k^{2} } =
\epsilon \mu -1 \; .
\label{5.2.97}
\end{eqnarray}

\noindent
Therefore, equations   (\ref{5.2.95b})  become
\begin{eqnarray}
H_{ab} = \Delta_{abmn} \; F^{mn} \; , \;\;
\Delta_{abmn} =  \epsilon_{0}\epsilon k^{2}
[  \; g_{am} + (\epsilon \mu -1) \;  u_{a}u_{m} \;  ] \;
[  \; g_{bn} + (\epsilon \mu -1)\;  u_{b}u_{n}  \;   ] \; . \;\;\;
\label{5.2.98}
\end{eqnarray}

\noindent
Just this representation for 4-rank tensor  relating
$H^{ab}$ to $F^{ab}$  in a uniform media  is given in the review [16].
A fresh review  of the history of  different electrodynamics constitutive equations
is given in recent work [19].

Now we are ready to introduce potentials.
The Maxwell equations are
\begin{eqnarray}
\partial _{a}  \; H^{ab}    =  j^{b}\; ,  \;\;
\partial _{a}  \;  ( \Delta^{abmn} F_{mn}  )     =  j^{b}\; ,
\label{5.2.99a}
\\
\partial_{c}\;  F_{ab} +  \partial_{a}\;  F_{bc} + \partial_{b}\;  F_{ca}  = 0 \; .
\label{5.2.99b}
\end{eqnarray}

\noindent
Potentials  $A_{b}$ are defined in such a way that  equations  (\ref{5.2.99b}) turn to identities:
\begin{eqnarray}
F_{ab} = \partial_{a} A_{b} - \partial_{b} A_{a} \; ;
\label{5.2.99c}
\end{eqnarray}

\noindent
With the help of (\ref{5.2.94b}),  the  $H^{ab}$ may be rewritten  as
\begin{eqnarray}
H^{ab} =
\epsilon_{0}\epsilon k^{2} \; (\partial^{a} A^{b} - \partial^{b} A^{a}) +
\epsilon_{0}\epsilon  \; (k^{2} -1)
( \; u^{a} \partial^{b}  - u^{b} \partial^{a}    \; ) \; ( u^{n} A_{n}) \; .
\label{5.2.99d}
\end{eqnarray}

\noindent
Therefore, eq. (\ref{5.2.99a}) leads us to
\begin{eqnarray}
\epsilon_{0}\epsilon k^{2} \; \partial _{a} (\partial^{a} A^{b} - \partial^{b} A^{a})
 +
\epsilon_{0}\epsilon   (k^{2} -1)  \partial _{a}
(  u^{a} \partial^{b}  - u^{b} \partial^{a}     )  ( u^{n} A_{n} ) =  j^{b}  . .
\label{5.2.100a}
\end{eqnarray}

\noindent
This is the main equation for electromagnetic 4-potentials in a media,
this equation is invariant under the ordinary Lorentz transformations
based on the vacuum light velocity.
For purely vacuum case, the factor  $(k^{2}-1)$  equals to zero and  (\ref{5.2.100a})
 takes the more familiar form
\begin{eqnarray}
 \partial _{a} \partial^{a} A^{b} - \partial^{b} (\partial _{a}  A^{a}) =
  \mu_{0} \; j^{b} \; .
\label{5.2.100b}
\end{eqnarray}

The scheme with modified Lorentz symmetry,  after
transition to 4-potential  $f_{CB} = \partial_{C} A_{B} -
\partial_{B} A_{C} $ leads to the simple wave equation (compare with (\ref{5.2.100a}))

\begin{eqnarray}
\partial^{B} \partial_{B} \; A^{C}  -  \partial_{C} \partial_{B} A^{B}=   j^{C}  \; .
\end{eqnarray}

No additional 4-velocity parameter enters this equation, so this
form of the  electrodynamics presumes a relative nature of the
mechanical motion;  also this equation describes waves propagating
in  space with the light velocity $kc$, which is  invariant  under
modified Lorentz formulas. In connection with these two
theoretical schemes,  a   point  of principle  must be stressed:
it might seem well-taken the requirement to perform
Poincar\'{e}-Einstein clock synchronization in the
 uniform  medias
with the help of  real light signals influenced by the  media,
which  leads us to the modified Lorentz symmetry.

\section{Discussion}

From formal mathematical  standpoint the all situation with the Maxwell's theory in uniform media
and its properties under the mechanical  motion of the inertial reference  frames looks
rather peculiar and designing. Two alternative possibilities exist.

I.

One may start with usual Maxwell's equation in terms of four electromagnetic vectors
${\bf E}, {\bf D}, {\bf B}, {\bf H}$ with  two additional restrain conditions
${\bf D} = \epsilon_{0} \epsilon {\bf E}$ and ${\bf B} = \mu_{0}\mu {\bf H}$,
then to translate the Maxwell equations in term of new electromagnetic  variables
${\bf d}, {\bf h}$ without any additional restrain conditions. It is mathematically
correct procedure evidently without any serious objection. Such a  form  of the
Maxwell's theory is remarkable because it  allows us to reveal the existence of
very simple symmetry in this  theory; namely, it is invariant with respect to the group of
modified Lorentz  transformations, based on the use of the speed of light $kc$
 in the  media instead of  the speed of light in the  vacuum applied in the  conventional
Lorentz  symmetry. This mathematical  result  might to be seen as
 an ideal realization of  the relativity principle in the  presence of the uniform media.
One might expect that  just such a symmetry for  electrodynamics in the uniform media
must be taken as a base for describing the properties
of all electromagnetic quantities under the  mechanical motion of the reference frame.

II.

The second  theoretical possibility is realized in the
Minkowski  approach to electrodynamic in uniform media.
In this approach only conventional Lorentz symmetry with vacuum speed of light is used.
Maxwell equation are formulated in covariant form
at the rest reference frame tied with the media,
in terms of two electromagnetic tensors $F^{ab}$ and $H^{ab}$. At the Lorentz translation
they transform independently, the Maxwell's equations are Lorentz invariant,
 but the additional restrain conditions  change their form.
To reach the formal relativistic invariance of the theory, the new modified form of the above
restrain condition by definition is taken as genuine and true. In virtue of the
construction procedure itself  this new form of the restrain  conditions is
automatically  Lorentz invariant.
At this  we must understand that  almost any equation,
through the  mathematical trick of that type,
 can be translated to some new form that will be   formally  Lorentz invariant.
 In this connection the question may be posed -- what is the new knowledge gained here.
Therefore, the  formal relativistical invariance is achieved in the Minkowski electrodynamics,
but his equations contain explicitly an additional physical parameter -- 4-velocity
of the moving media (or differently, of the reference frame).

This means, that electrodynamics by Minkowski  in the media
presumes the absolute nature of the mechanical motion.
In this  connection, we  might  recall that  Special relativity theory had
started many years ago somehow from the
requirement -- to  mechanical motion in electrodynamics should  be a relative concept,
no physical experiment is  able to reveal the inertial motion of
the reference frame.
In a sense, the time has played  a joke with our previous, and  being taken seriously,
 theoretical arguments about
 symmetry of electromagnetic equations and  relativity principle.
The special relativity had been constructed to avoid an absolute velocity concept,
but this absolute velocity  concept again has returned in the frame of electrodynamics
by Minkowski.

In any case   simply ignoring the existence of  the modified Lorentz
symmetry  in the Maxwell theory is not a correct
 attitude. It is hard to believe that this modified
 symmetry for the   Maxwell's theory in the uniform media is of no physical meaning
 and value.

\end{document}